\documentclass{nature}

\newcommand{\msun}{M_{\odot}}
\newcommand{\Mvir}{M_{\rm vir}}
\newcommand{\avir}{\alpha_{\rm vir}}
\newcommand{\tdyn}{t_{\rm dyn}}
\newcommand{\phibh}{\phi_{\rm BH}}

\newcommand{\fm}{f_{\rm m}}
\newcommand{\fmbh}{f_{\rm m-BH}}
\newcommand{\fmcap}{f_{\rm m-cap}}
\newcommand{\Rco}{R_{\rm co}}
\newcommand{\Mco}{M_{\rm co}}
\newcommand{\phico}{\phi_{\rm co}}

\newcommand{\ltsim}{\protect\raisebox{-0.5ex}{$\:\stackrel{\textstyle <}
	{\sim}\:$}}

\title{Stars Form By Gravitational Collapse, Not Competitive Accretion}

\author{Mark R. Krumholz$^1$, Christopher F. McKee$^{2,3}$, \& Richard
I. Klein$^{3,4}$}

\begin{document}

\maketitle

\begin{affiliations}
\item Hubble Fellow; Astrophysics Department, Princeton University,
Princeton, NJ 08544
\item Physics Department, UC Berkeley, Berkeley, CA 94720
\item Astronomy Department, UC Berkeley, Berkeley, CA 94720
\item Lawrence Livermore National Laboratory, Livermore, CA 94550
\end{affiliations}

\begin{abstract}
There are now two dominant models of how stars form: gravitational
collapse theory holds that star-forming molecular clumps, typically
hundreds to thousands of $\msun$ in mass, fragment into gaseous cores
that subsequently collapse to make individual stars or small multiple
systems\cite{shu87,padoan02,larson05}. In contrast, competitive accretion
theory suggests that at birth all stars are much smaller than the
typical stellar mass ($\sim 0.5$ $\msun$), and that final stellar
masses are determined by the subsequent accretion of unbound gas from
the clump\cite{bonnell98,bonnell01a,bonnell01b,bonnell04,bate05}.
Competitive accretion models explain brown dwarfs and free-floating
planets as protostars ejected from star-forming clumps before
accreting much mass, predicting
that they should lack disks, have high velocity dispersions, and form
more frequently in denser clumps\cite{bate02a,bate03,mohanty05}. They
also predict that mean stellar mass should vary within the
Galaxy\cite{bate05}. Here we derive a simple estimate for the rate of
competitive accretion as a function of the star-forming environment,
based partly on simulations\cite{krumholz05d}, and determine in what
types of environments competitive accretion can occur. We show that no
observed star-forming region produces significant competitive
accretion, and that simulations that show competitive
accretion do so because their properties differ from those
determined by observation. Our result shows that stars form by
gravitational collapse, and explains why observations have failed to
confirm predictions of the competitive accretion scenario.
\end{abstract}

In both theories, a star initially forms when a gravitationally bound
gas core collapses. The crucial distinction between them
is their prediction for what happens
subsequently. In gravitational collapse, after a protostar has
consumed or expelled all the gas in its initial core, it may continue
accreting from its parent clump. However, it will not accrete
enough to substantially change its mass\cite{mckee03,padoan05}.
In contrast, competitive accretion requires that the amount accreted
after the initial core is consumed be substantially larger than the
protostellar mass. We define $\fm\equiv \dot{m}_* \tdyn/m_*$ as the
fractional change in mass that a protostar of mass $m_*$ undergoes
each dynamical time $t_{\rm dyn}$ of its parent clump, starting after
the initial core has been consumed. Gravitational collapse holds that
$\fm \ll 1$, while competitive accretion requires $\fm\gg 1$.

Consider a protostar embedded in a molecular clump of
mass $M$ and mass-weighted one-dimensional velocity dispersion
$\sigma$. Competitive accretion theories usually begin with seed
protostars of mass $m_*\approx 0.1$
$\msun$\cite{bonnell98,bonnell01a,bonnell01b,bonnell04}, 
so we adopt this as a typical value. We consider two possible
geometries: spherical clumps of radius $R$ and filaments of radius $R$
and length $L$, $L\gg R$. These extremes bracket real star-forming
clumps, which have a range of aspect ratios. The virial mass for
(spherical, filamentary) clumps is 
\begin{equation}
\label{mvirdef}
\Mvir \equiv \left(\frac{5\sigma^2 R}{G}, \frac{2\sigma^2
L}{G}\right),
\end{equation}
and the virial parameter is $\avir \equiv
\Mvir/M$\cite{bertoldi92,fiege00b}.
The dynamical time is $\tdyn\equiv R/\sigma$.

First suppose that the gas that the protostar is accreting is not
collected into bound structures on scales smaller than the entire
clump. Since the gas is unbound, we may neglect its
self-gravity and treat this as a problem of a non-self-gravitating gas
accreting onto a point particle. This process is Bondi-Hoyle accretion
in a turbulent medium, which gives an accretion rate\cite{krumholz05d}
\begin{equation}
\label{mdotbhcompacc}
\dot{m}_* \approx 4\pi \phibh \overline{\rho} \frac{(G
m_*)^2}{(\sqrt{3}\sigma)^3},
\end{equation}
where $\overline{\rho}$ is the mean density in the clump. The factor 
$\phibh$ represents the effects of turbulence, which we
estimate in terms of $\sigma$, $m_*$ and $R$ in the Supplementary
Information.\cite{krumholz05d}
From (\ref{mdotbhcompacc}) and the definitions of the virial parameter
and the dynamical time, we find that accretion of unbound gas gives
\begin{equation}
\label{fracmassgain}
\fmbh =
\left(14.4, 3.08\frac{L}{R}\right) \phibh
\avir^{-2} \left(\frac{m_*}{M}\right)
\end{equation}
for a (spherical, filamentary) star-forming region. From this result,
we can immediately see that competitive accretion is most effective
in low mass clumps with virial parameters much smaller than unity.

Table \ref{regionres} shows a broad sample of observed star-forming
regions. None of them have a value of $\fmbh$ near unity, inconsistent
with competitive accretion and in agreement with gravitational
collapse. Note that the Bondi-Hoyle rate is an upper limit on the
accretion. If stars are sufficiently close-packed, their tidal radii
will be smaller than their Bondi-Hoyle radii, and the accretion rate
will be lower\cite{bonnell01a}. Also note that radiation pressure will
halt Bondi-Hoyle accretion onto stars larger than $\sim 10$
$\msun$\cite{edgar04}.

The second possible way that a star could gain mass is by capturing
and accreting other gravitationally bound cores. We can analyze this
process by some simple approximations. First, when a star collides
with a core it begins accreting gas from it, causing a drag
force\cite{ruffert94b}. If drag dissipates enough energy, the two
become bound. We can therefore compute a critical velocity below which
any collision will lead to a capture and above which it will
not. Second, cores and stars should inherit the velocity dispersion of
the gas from which they form, so we assume they have Maxwellian
velocity distributions with dispersion $\sigma$. The true functional
form may be different, but this will only affect our estimates by a
factor of order unity. Third, we neglect the range of core sizes, and
assume that all cores have a generic radius $\Rco$ and mass
$\Mco$. Competitive accretion requires $\Mco \leq m_*$, so we take
$\Mco = m_*$, which gives the highest possible capture rate. Finally,
we make use of an
important observational result: cores within a molecular clump have
roughly the same surface density as the clump itself\cite{larson81},
$\Sigma=M(\pi R^2, 2 R L)^{-1}$ for (spherical, filamentary)
clumps. This enables us to compute the escape velocity from the
surface of a core in terms of the properties of the clump,
\begin{equation}
v_{\rm esc} = \left[
\left(10,
\sqrt{8\pi\frac{L}{R}}\right) \avir^{-1} \sqrt{\frac{\Mco}{M}}
\right]^{1/2} \sigma.
\end{equation}

With these approximations, it is straightforward to compute the amount
of mass that a protostar can expect to gain by capturing other
cores. In the Supplementary Information, we show it is
\begin{equation}
\label{fmcapeqn}
\fmcap = \left(0.42, 0.36\right)
\phico 
\left[4+2 u_{\rm esc}^2 - 
\left(4+7.32\, u_{\rm esc}^2\right)
\exp\left(-1.33\,u_{\rm esc}^2\right)\right],
\end{equation}
where $\phico$ is the fraction of the parent clump mass that is in
bound cores and $u_{\rm esc}\equiv v_{\rm esc}/\sigma$. Surveys
generally find core mass fractions of $\phico \sim
0.1$\cite{motte98,testi98,johnstone01}, so we adopt this as a typical
value, giving the numerical values of $\fmcap$ shown in Table
\ref{regionres}. As with $\fmbh$, all the estimated values are well
below unity.

If we let $\fm=\fmbh+\fmcap$, then we can use our simple models to
determine where in parameter space a star-forming clump must fall to
have $\fm\ge 1$. For simplicity, consider a spherical clump with
fixed $\phibh=5$ and $\phico=0.1$ (typical values for observed
regions), and a seed protostar of mass $m_*=0.1$. In this case, both
$\fmbh$ and $\fmcap$ are functions of 
$\avir^2 M$ alone, and we find $\fm\ge 1$ for $\avir^2 M < 8.4$
$\msun$. The functional dependence is more complex if we include
filamentary regions and allow $\phibh$ and $\phico$ to vary,
but the qualitative result is unchanged.
Observed star-forming regions have $\avir\approx 1$ and $M\approx
10^2-10^4$ $\msun$\cite{plume97}, which produces $\fm\ll 1$. No known
star-forming region has $\avir^2 M$ small enough for competitive
accretion to work. Thus, the cores from which stars form must contain
all the mass they will ever have, which is the gravitational
collapse model.

Our simple estimate of $\fm$ is consistent with simulations of
competitive accretion as well, and explains why competitive accretion
works in the simulations. All competitive accretion simulations have
virial parameters $\avir\ll 1$. In some cases the simulations
start in this
condition\cite{bonnell01a,bonnell01b,klessen00a,klessen01}, with
$\avir\approx 0.01$ as a typical choice. In other
cases, the virial parameter is initially of order unity, but as
turbulence decays in the simulation it decreases to $\ll 1$ in roughly
a crossing time\cite{bate02a,bate02b,bate03,bonnell04}. Once
competitive accretion gets going in these simulations, they have
reached $\avir\ll 1$ as well. In addition, many of the simulations
consider star-forming clumps of masses considerably smaller than the
$\sim 5000$ $\msun$ typical of most galactic star
formation\cite{plume97}, with $M\ltsim 100$ $\msun$ not
uncommon. Consequently, the simulations have $\avir^2 M \ltsim 10$
$\msun$, which explains why they find competitive accretion to be
important. Note that simulations where turbulence decays will have
$\phibh\approx 1$, rather than the typical value of $\phibh=5$ we have
used for real regions, but this does not substantially modify our
conclusions.

Three other aspects of the simulations even further increase their
estimate of $\fm$. First, the Bondi-Hoyle radius of a $0.1$ $\msun$
seed protostar in a typical clump is only 5 AU, a smaller scale than
any of the competitive accretion simulations resolve. This
under-resolution may enhance accretion\cite{krumholz05d}. Second,
small virial parameters lead most of
the mass to collapse to stars, giving $\phico\sim 0.5-1$ after a
dynamical time, and also tend to make the cloud fragment into smaller
pieces, lowering $M$. Third, rapid collapse leaves no time for
large cores to assemble. For example, one simulation of a
$\sim 1000$ $\msun$ clump produces no cores larger than $1$
$\msun$\cite{bonnell04}, inconsistent with
observations that find numerous cores more massive than
this in similar regions\cite{johnstone01,beuther04b}. With no large
cores, large stars can only form via competitive accretion.

Thus, our results are consistent with the simulations, 
but they show that the simulations are not modelling
realistic star-forming clumps. One might argue that all clumps do
enter a phase with $\avir \ll 1$ that occurs rapidly and has therefore
never been observed, but that most stars are formed during this
collapse phase. In this scenario, though, protostars associated with
observed star-forming regions should have systematically lower masses
than the field star population, since they were formed before the
collapse phase when competitive accretion might work. One would also
expect to see a systematic variation in mean stellar mass with age in
young clusters, corresponding to cluster evolution into a state more
and more favorable to competitive accretion. This is not observed.

We hypothesize that the primary problem with the simulations, the
reason they evolve to $\avir\ll 1$, is that they omit feedback from
star formation. Recent observations of protostellar outflow
cavities show that outflows inject enough energy to sustain the
turbulence and prevent the virial parameter from declining to values
much less than unity\cite{quillen05}. Another possible problem in the
simulations is that they simulate isolated clumps containing too
little material. Real clumps are embedded in molecular clouds, and
large-scale turbulent motions in the clouds may cascade down to the
clump scale and prevent the turbulence from decaying. A third
possibility is that turbulence decays too quickly in the simulations
because they do not include magnetic fields and their initial
velocity fields, unlike in real clumps, are balanced rather than
imbalanced between left- and right-propogating modes\cite{cho03}.

One implication of our work is that brown dwarfs need not have
been ejected from their natal clump, so their velocity dispersions
should be at most slightly greater than those of stars, and their
frequency need not change as a
function of clump density. This also removes a discrepancy between
observations showing that brown dwarfs have disks\cite{mohanty05} and
theoretical models of their origins. Another conclusion is that the
mean stellar mass need not vary from one star-forming region to
another as competitive accretion predicts, removing a discrepancy
between theory\cite{bate05} and observations that have thus far failed
to find any substantial variation in typical stellar mass with
star-forming environment. In the gravitational collapse scenario, the
mean stellar mass may be roughly constant in the Galaxy, but may vary
with the background radiation field in starburst regions and in the
early universe\cite{larson05}.

\bibliography{ms}

\begin{addendum}
\item[Supplementary Information]is linked
to the online version of the paper at \\www.nature.com/nature.
\item[Acknowledgements] We thank R.~T. Fisher for discussions and
P. Padoan for comments. This work
was supported by grants from NASA through the Hubble Fellowship, GSRP,
and ATP programs, by the NSF, and by the US DOE
through Lawrence Livermore National Laboratory. Computer simulations
for this work were performed at the San Diego Supercomputer
Center (supported by the NSF), the National Energy Research Scientific
Computer Center (supported by the US DOE), and Lawrence Livermore
National Laboratory (supported by the US DOE).
\item[Author Information] Reprints and Permissions information
is available at \\npg.nature.com/reprintsandpermissions. Authors declare
they have no competing financial interests. Correspondence and
requests for materials should be addressed to
M.R.K. \\(krumholz@astro.princeton.edu).
\end{addendum}

\clearpage

\begin{table}
\begin{tabular}{ccccccc}
Name &
Mass &
Type &
$M$ ($\msun$) &
$R$ (pc) &
$L$ (pc) &
$\sigma$ (km s$^{-1}$)
\\
L1495 I\cite{kramer91} & Low & Sph. & 410 & 2.1 & - & 0.58 \\
L1495 II\cite{kramer91} & Low & Sph. & 950 & 2.4 & - & 0.67 \\
L1709\cite{fiege00b} & Low & Fil. & 140 & 0.23 & 3.6 & 0.48 \\
L1755\cite{fiege00b} & Low & Fil. & 171 & 0.15 & 6.3 & 0.53 \\
W44\cite{plume97} & High & Sph. & 16000 & 0.35 & - & 3.9 \\
W75(OH)\cite{plume97} & High & Sph. & 5600 & 0.25 & - & 3.5 \\
$\int$-fil\cite{fiege00b} & High & Fil. & 5000 & 0.25 & 13 & 1.41 \\
N-fil\cite{fiege00b} & High & Fil. & 16000 & 2.3 & 88 & 1.54
\end{tabular}
\caption{\label{regiontab}
\textbf{Sample star-forming regions.}
Sph. = spherical, Fil. = Filamentary, $\int$-fil = Orion integral
filament, N-fil = Orion North filament. For L1495 I and II, the data
are from Kramer \& Winnewisser's $^{12}$CO observations, and the masses
are Kramer \& Winnewisser's $M_{\rm CO}$. For W44 and W75(OH), the
data are from Plume \textit{et al.}'s CS $J=5\rightarrow 4$
observations, and the masses are Plume \textit{et al.}'s $M_n$. Note
that, since CS $J=5\rightarrow 4$ is a very high density tracer, it
biases the results to small virial parameters by excluding low
density parts of the clump.
}
\end{table}

\clearpage

\begin{table}
\begin{tabular}{ccccccc}
Name &
$\avir$ &
$\phibh$\cite{krumholz05d} &
$\fmbh$ &
$u_{\rm esc}$ &
$\fmcap$
\\
L1495 I    & 2.0 &  2.4 & 0.0022 & 0.28 & 0.0015\\
L1495 II   & 1.3 &  3.0 & 0.0026 & 0.28 & 0.0015\\
L1709      & 2.8 &  0.93 & 0.0042 & 0.44 & 0.0072\\
L1755      & 4.8 &  0.54 & 0.0017 & 0.40 & 0.0052\\
W44        & 0.39 & 6.4 & 0.0038 & 0.25 & 0.0010\\
W75(OH)    & 0.63 & 5.2 & 0.0034 & 0.26 & 0.0011\\
$\int$-fil & 2.4 &  4.1 & 0.0023 & 0.26 & 0.00097\\
N-fil	   & 6.2 &  5.7 & 0.0001 & 0.11 & 0.00003
\end{tabular}
\caption{\label{regionres}
\textbf{Computed properties of sample star-forming regions.}
$\int$-fil = Orion integral filament, N-fil = Orion North filament.
}
\end{table}

\end{document}